# Switching-Algebraic Calculation of Banzhaf Voting Indices

[1]Ali Muhammad Rushdi and [2,3]Muhammad Ali Rushdi

[1]Department of Electrical and Computer Engineering, Faculty of Engineering,
King Abdulaziz University, P. O. Box 80200, Jeddah, 21589, Kingdom of Saudi Arabia
{arushdi@kau.edu.sa; arushdi@ieee.org}

[2]Department of Biomedical and Systems Engineering, Faculty of Engineering,
Cairo University, Giza 12613, Arab Republic of Egypt
{mrushdi@eng1.cu.edu.eg}
[3]School of Information Technology,
New Giza University, Giza 12256, Arab Republic of Egypt
{Muhammad.Rushdi@ngu.edu.eg}

**Abstract**: This paper employs switching-algebraic techniques for the calculation of a fundamental index of voting powers, namely, the total Banzhaf power. This calculation involves two distinct operations: (a) Boolean differencing or differentiation, and (b) computation of the weight (the number of true vectors or minterms) of a switching function. Both operations can be considerably simplified and facilitated if the pertinent switching function is symmetric or it is expressed in a disjoint sum-of-products form. We provide a tutorial exposition on how to implement these two operations, with a stress on situations in which partial symmetry is observed among certain subsets of a set of arguments. We introduce novel Boolean-based symmetry-aware techniques for computing the Banzhaf index by way of two prominent voting systems. These are scalar systems involving six variables and nine variables, respectively. The paper is a part of our on-going effort for transforming the methodologies and concepts of voting systems to the switching-algebraic domain, and subsequently utilizing switching-algebraic tools in the calculation of pertinent quantities in voting theory.

## 1. Introduction

Weighted voting systems constitute a major class of yes-no systems, in which a weight is assigned to each voter, and a suitable quota or threshold is selected. A bill (proposal, resolution, or amendment) is accepted (passed) if the sum of weighted votes in favor of it reaches or goes beyond the selected threshold [1-3]. The voting power of an individual voter $V$ in a voting system is the probability that this specific voter is decisive, which is decided by the number of ways the voter can bring about a swing in the outcome, and ultimately by the rule for aggregating votes into a single two-valued Boolean outcome $f(X)$. Here the binary vector $X = [X_1 \ X_2 \ ... \ X_n]^T$ is an n-tuple of the binary votes $X_i$ ( $1 \leq i \leq n$ ) expressed by the voters, where $X_i$ is 1 or 0 if voter $i$ is among the proponents (a yes-voter) or among the opponents (a no-voter), respectively. Many indices of voting power are in use nowadays. Two of these are effectively the mainstream standard ones. These are the

Banzhaf index [4-7] (also occasionally referred to as the Penrose–Banzhaf index (See [8]), the Banzhaf–Coleman index (See [9]) or the Penrose–Banzhaf–Coleman index), and the Shapley–Shubik power index [10]. These two indices are based on coalitional and permutational considerations, respectively. Usually, a change in the voting scheme that increases the power of a voter on one particular index tends to increase this power on the other index as well, and vice versa.

Our main concern in this paper is the switching-algebraic computation of the Banzhaf index or Banzhaf voting power of a voter $V$ in a weighted voting system, which is typically called the total Banzhaf power $TBP(V)$ of the individual voter $V$ [2]. It is the number of times voter $V$ is decisive, i.e., the number of winning states or configurations in which this voter is among the proponents (yes-voters) of the debated proposal such that a switch of the voter to join the opponents (no-voters) changes the system state from one of winning (proposal approval) to that of losing (proposal rejection). Rushdi and Ba-Rukab [7] coined the name of a "primitive coalition" for each of the $2^n$ states or configurations for an n-member weighted voting system. The Total Banzhaf Power $TBP(V)$ is then the number of primitive winning coalitions ($PWCs$) of which voter $V$ is a member such that when $V$ switches sides (from a proponent to an opponent), the coalition swings to a primitive losing coalition ($PLC$). In the sequel, we employ the typical assumption that the variables $X_i$ ( $1 \leq i \leq n$) are statistically independent. This assumption is basically associated with the supposition that the primitive coalitions, states or configurations are equally probable.

The raw value for the total Banzhaf power of voter number $m$ has the following switching-algebraic definition [6, 7]

$$TBP(X_m) = wt\left(\frac{\partial f(X)}{\partial X_m}\right), \quad (1 \leq m \leq n). \quad (1)$$

Here, the symbol $\frac{\partial f(X)}{\partial X_m}$ denotes the partial derivative of the voting system Boolean function $f(X)$ w.r.t. its argument $X_m$ (See Appendix A), while the symbol $wt(...)$ denotes the weight or number of true vectors of a switching function (See Appendix B). The most appealing normalization of the aforementioned raw value $TBP(X_m)$ is obtained through dividing the raw value by the sum of all such values. This yields the following normalized total Banzhaf power:

$$NTBP(X_m) = TBP(X_m) / \sum_{k=1}^{n} TBP(X_k), \quad (1 \leq m \leq n). \quad (2)$$

The resulting normalized powers are now situated within the real unit interval [0.0, 1.0], which allows a probability interpretation for them, and also facilitates comparison with other types of voting powers such as the Shapley-Shubik indices [11]. There are several algorithms for calculating the Banzhaf power index that usually employ recursion and problem decomposition into sub-problems, e.g., techniques of dynamic programming [12], enumeration methods [12], and generating-function methods [13, 14]. There are also other algorithms for approximating this index via random sampling and Monte Carlo simulations [15, 16]. The technique to be proposed herein is a switching-algebraic technique that is basically an enumeration method. This technique tries to make the most of recent developments in switching theory as well as of symmetry features that are inherent to many weighted voting systems.

The topic of this paper is related to certain topics within reliability theory in a few ways.

(a) First, we note that the Banzhaf index and other voting indices are sometimes utilized as importance measures in reliability theory, and there is a striking similarity between the Banzhaf index and the Birnbaum importance measure of reliability systems [17-22].
(b) Moreover, our earlier explorations of the reliability of threshold systems [23] indicates the existence of a handy methodology for the study of weighted voting systems, namely the methodology set by the theory of threshold switching functions [23-26]. This theory has already matured within digital-design circles, and can be fruitfully utilized in, and appropriately adapted to, the study of both threshold reliability systems and weighted voting systems.
(c) Finally, the computation of the Banzhaf voting index involves two distinct operations: (a) Boolean differencing or differentiation [6, 7, 25], and (b) computation of the weight (the number of true vectors or minterms) of a switching function [24-26]. We simplify and facilitate both operations by expressing the pertinent switching function in a disjoint sum-of-products form [27-32] by borrowing techniques of disjointness from the reliability literature.

The remainder of this paper is structured as follows. Section 2 explains some of the basic concepts and nomenclature. Section 3 explores Banzhaf indices for symmetric switching functions. Section 4 introduces novel Boolean-based symmetry-aware techniques for computing the Banzhaf index by way of two prominent voting systems. These are scalar systems describing 6-variable and 9-variable versions of the European Economic Community. To make the paper self-contained, we supplement and support its main text with two appendices. Appendix A introduces the concept and calculus of the Boolean difference (Boolean derivative). Appendix B discusses the concept and properties of the weight of a switching function, and then highlights a variety of methods and shortcuts for computing it.

## 2. Basic Concepts and Nomenclature

**A yes-no voting system**: A voting system which offers a choice between adopting a potentially forthcoming alternative (an amendment, a resolution or a bill) versus the status quo, which stands as an already existing alternative [2]. This system is described herein by a switching (two-valued Boolean) threshold function $f(X)$, such that $f(X) = 1$ if the resolution considered is passed and $f(X) = 0$ if the resolution is rejected. The function $f(X)$ is similar to the success function of a coherent threshold reliability system [23, 32]. Here the binary vector $X = [X_1 \ X_2 \ ... \ X_n]^T$ is an n-tuple of the votes $X_i$ [$1 \leq i \leq n$] cast by voters, where the value of $X_i$ is an indicator whether voter $i$ approves ($X_i = 1$) or disapproves ($X_i = 0$) the debated resolution.

**A coalition**: Any set comprised solely of yes-voters in a yes-no voting system. The coalition is winning if the disputed alternative in question is upheld (i.e., if the specific amendment, resolution or bill considered is passed), and otherwise the coalition is losing. The two extreme cases for a coalition is the empty coalition, to which no voter belongs, and the grand coalition, to which all voters belong [2, 33].

**A primitive coalition**: A specification of the status or configuration of all voters (possibly including both types of voters (yes-voters and no-voters)) [7]. This can be a primitive winning coalition (PWC), that corresponds to a true vector (minterm) of $f(X)$, or a primitive losing coalition (PLC), that corresponds to a false vector (minterm) of $f(X)$. The concept of a primitive coalition coincides with those of a line of a truth table or a cell of a Karnaugh map of $f(X)$. Two prominent primitive coalitions are the all-0 primitive coalition (which coincides with the empty coalition), and the all-1 primitive coalition (which coincides with the grand coalition). The concept of a primitive coalition is quite convenient for Boolean-based analysis, but it is admittedly alien to voting theory. A primitive coalition is typically a mixture of yes-voters and no-voters (typically expressed by a product of uncomplemented literals and complemented ones). By contrast, the concept of a coalition is popular in voting theory. It concerns solely yes-voters (and hence it is expressed by a product of uncomplemented literals only).

**A dummy voter**: A voter $P$ who has no say in the outcome of the voting system, since $TBP(P)$ is strictly equal to 0. This voter has no power whatsoever, since he or she cannot influence the passing of a resolution in any case. The existence of a dummy voter defeats the purpose of the voting system, which should allow each individual voter some plausible chance, however small, to affect or influence the decisions made by the system. If the voting system has a dictator or a dictating clique then the remaining voters are definitely all dummies. The non-permanent members of the United Nations security council (UNSC) are not dummies in the technical sense of the word, albeit they are definitely almost dummies. The $TBP$ of a non-permanent member is alarmingly negligible compared to that of a permanent member but it is not strictly equal to 0. The $TBP$ of all ten non-permanent members of the UNSC put together is slightly less than that of a single permanent member. The voting power in the UNSC is divided into six shares, with five of them divided evenly among the permanent members, and with the sixth share split into sub-shares circulating among alternating representatives of the rest of the world.

**A symmetric switching function (SSF)**: A two-valued Boolean function depicted [32, 34] as

$$f(X) = Sy(n; A; X) = Sy(n; \{a_1, a_2, \ldots, a_m\}; X_1, X_2, \ldots, X_n). \qquad (3)$$

The SSF in (3) is completely characterized by its number of inputs $n$, its inputs $X = [X_1, X_2, \ldots, X_n]^T$, and its characteristic set

$$A = \{a_0, a_1, \ldots, a_m\} \subseteq I_{n+1} = \{0, 1, 2, \ldots, n\}, \{m \leq n\}. \qquad (4)$$

This function has the value 1 if and only if the arithmetic sum $\sum_{j=1}^{n} X_j$ belongs to the characteristic set $A$, and has the value 0, otherwise. Symmetry is preserved by each of the unary operation *NOT* and binary operations *AND*, *OR*, and *XOR*. Specifically, the complement $\bar{f}$ of the SSF in (3) is also symmetric, and it possesses a characteristic set $\bar{A}$ that complements the original characteristic set $A$ w.r.t. the universe $I_{n+1} = \{0, 1, 2, \ldots, n\}$. The complementary set $\bar{A}$ is given by the set difference $(I_{n+1}/A)$, also denoted as $(I_{n+1} - A)$, or

$$\bar{A} = \{0, 1, 2, \ldots, n\} - \{a_0, a_1, \ldots, a_m\}, \qquad (5)$$

and hence, $\bar{f}$ can be expressed as

$$\bar{f}(X) = Sy(n; \bar{A}; X). \qquad (6)$$

Moreover, the *ANDing*, *ORing*, and *XORing* of two SSFs $Sy(n; A_1; X)$ and $Sy(n; A_2; X)$ (which share the same arguments $X$, and are of characteristic sets $A_1$ and $A_2$, respectively) results in SSFs whose characteristic sets are equal to the intersection, union, and XORing of the original sets $A_1$ and $A_2$, respectively, *i.e.*,

$$Sy(n; A_1; X) \land Sy(n; A_2; X) = Sy(n; A_1 \cap A_2; X), \tag{7}$$

$$Sy(n; A_1; X) \lor Sy(n; A_2; X) = Sy(n; A_1 \cup A_2; X), \tag{8}$$

$$Sy(n; A_1; X) \oplus Sy(n; A_2; X) = Sy(n; A_1 \oplus A_2; X). \tag{9}$$

**A threshold switching function**: A switching function $f(X)$ of $n$ variables characterized by $(n + 1)$ (rather than $2^n$) coefficients, namely a threshold $T$ and weights $W = [W_1, W_2, \ldots, W_n]^T$, such that [23]

$$f(X) = 1 \quad \text{iff} \quad F(X) = \sum_{i=1}^{n} W_i X_i \geq T. \tag{10}$$

A threshold switching function might be described as scale-invariant, since multiplying every weight and the threshold by the same positive constant does not change the function. A weighted voting system characterized by $f(X)$ is typically denoted by $(T; W_1, W_2, \ldots, W_n)$.

**A semi-coherent switching function:** A switching function $f(X)$ possessing the property of monotonicity, i.e., it is a monotonically non-decreasing function. Since monotonicity implies causality (for non-constant functions representing non-fictitious systems), then a semi-coherent $f(X)$ possesses the property of causality as well.

**A (fully) coherent switching function:** A semi-coherent switching function $f(X)$ that additionally possesses the property of component relevancy (for all components). Such a coherent function $f(X)$ is a unate function of an all-positive polarity [26], that can have a sum-of-products representation consisting solely of uncomplemented literals. It has a unique and canonical minimal sum (disjunction of a minimal number of prime implicants that collectively cover it) that exactly equals its complete sum (disjunction of all prime implicants).

**A coherent threshold switching function**: A threshold switching function with strictly positive weights and threshold. It is used to describe the decision made by a scalar weighted voting system, or the success of a threshold reliability system [23], also called a weighted-k-out-of-n system. In particular, the success $S(k, n, X)$ of a $k$-out-of-$n$: G reliability system is a symmetric coherent threshold function with unit weights and a threshold equal to $k$ [23], since

$$\{S(k, n, X) = 1\} \quad iff \quad \{\sum_{i=1}^{n} X_i \geq k\}. \tag{11}$$

**Monotonicity**: For a switching function $f(X)$, monotonicity means that it is monotonically non-decreasing, i.e.,

$$f(X) \geq f(Y) \quad \text{for} \quad X \geq (Y), \tag{12a}$$

or equivalently

$$f(X | X_i = 1) \geq f(X | X_i = 0). \tag{12b}$$

If a set of yes-voters constitutes a winning coalition, then any superset of yes-voters is a winning coalition as well. If a set of yes-voters forms a losing coalition, then any subset of yes-voters is also a losing coalition.

**Causality**: For a switching function $f(\mathbf{X})$, causality means that it is 0 when its argument is the all-0 vector, and it is 1 when its argument is the all-1 vector, i.e.,

$$\{f(\mathbf{0}) = 0\} \wedge \{f(\mathbf{1}) = 1\}. \tag{13}$$

For a yes-no voting system, causality means that the empty and grand coalitions (corresponding to the all-0 and the all-1 primitive coalitions, respectively) are losing and winning ones, respectively.

**Component relevancy (for all components)**: For a switching function $f(\mathbf{X})$, component relevancy means that

$$\frac{\partial f(\mathbf{X})}{\partial X_i} = f(\mathbf{X}|X_i = 0) \oplus f(\mathbf{X}|X_i = 1) \neq 0 \quad \text{identically for } 1 \leq i \leq n, \tag{14a}$$

i.e., there exists at least one instance of $\mathbf{X}$ such that

$$f(\mathbf{X}|X_i = 0) \oplus f(\mathbf{X}|X_i = 1) \neq 0, \text{ for } 1 \leq i \leq n. \tag{14b}$$

For a yes-no voting system, component relevancy means that no voting member is dummy. This means that every voting member must belong to a winning coalition that becomes a losing one if that member alone defects from it. In other words, the defection of the member is *decisive* or *critical* for the winning status of the coalition.

### 3. Banzhaf Indices for Symmetric Switching Functions

A symmetric switching function $Sy(n; \mathbf{A}; \mathbf{X})$ is characterized by its Boole-Shannon expansion about any of its variables $X_m$ ($1 \leq m \leq n$). This expansion might be stated as follows [32]

$$Sy(n; \mathbf{A}; \mathbf{X}) = \bar{X}_m \, Sy(n-1; \mathbf{B}; \mathbf{X}/X_m) \vee X_m \, Sy(n-1; \mathbf{C}; \mathbf{X}/X_m), \ (1 \leq m \leq n), \tag{15}$$

where the two sets $\mathbf{B}$ and $\mathbf{C}$ are both subsets of the set the first $n$ non-negative integers $\mathbf{I}_n = \{0, 1, 2, \ldots, n-1\}$, and they are precisely defined as

$$\mathbf{B} = \mathbf{A} \cap \mathbf{I}_n, \tag{16}$$

$$\mathbf{D} = \{a_0 - 1, a_1 - 1, \ldots, a_m - 1\}, \tag{17}$$

$$\mathbf{C} = \mathbf{D} \cap \mathbf{I}_n. \tag{18}$$

The definitions of the two sets $\mathbf{B}$ and $\mathbf{C}$ might be restated as follows

$$\begin{aligned}
\mathbf{B} &= \mathbf{A} & \text{if } a_m \neq n, & \tag{19a}\\
\mathbf{B} &= \mathbf{A} - \{n\} & \text{if } a_m = n, & \tag{19b}
\end{aligned}$$

$$\begin{aligned}
\mathbf{C} &= \mathbf{D} & \text{if } a_0 \neq 0, & \tag{20a}\\
\mathbf{C} &= \mathbf{D} - \{-1\} & \text{if } a_0 = 0. & \tag{20b}
\end{aligned}$$

The Boole-Shannon expansion (15) is conveniently applicable within a specific region of useful validity, in which the characteristic set $A$ is a strict subset of the universal set $I_{n+1}$ and a strict superset of the empty set $\phi$. Within this region of useful validity, the expansion can be recursively applied till one of the following boundaries $(A = I_{n+1})$ or $(A = \phi)$ is reached. At the boundaries, the recursion is terminated by one of the boundary conditions

$$Sy(n; I_{n+1}; X) = 1, \qquad (21)$$

$$Sy(n; \phi; X) = 0, \qquad (22)$$

where $I_{n+1}$ is the set of the first $(n + 1)$ non-negative integers, and $\phi = \{\}$ is the set to which no element belongs.

The two terms in the R.H.S. of (15) is disjoint since $\bar{X}_m$ appears in the first term while $X_m$ appears in the second. Therefore, it is legitimate to replace the OR operator (∨) by an XOR operator (⊕) in (15), namely

$$Sy(n; A; X) = \bar{X}_m \, Sy(n - 1; B; X/X_m) \oplus X_m \, Sy(n - 1; C; X/X_m), \quad (1 \leq m \leq n), \qquad (23)$$

Hence, the Boolean derivative of the SSF $Sy(n; A; X)$ w.r.t. $X_m$ is readily obtained (see Appendix A) as another SSF given by

$$\frac{\partial Sy(n;A;X)}{\partial X_m} = Sy(n - 1; B; X/X_m) \oplus Sy(n - 1; C; X/X_m), \quad (1 \leq m \leq n), \qquad (24)$$

$$\frac{\partial Sy(n;A;X)}{\partial X_m} = = Sy(n - 1; B \oplus C; X/X_m), \quad (1 \leq m \leq n). \qquad (25)$$

The total Banzhaf power is given (according to (1), (25) and (B.13)) by

$$TBP(X_m) = wt\left(\frac{\partial Sy(n;A;X)}{\partial X_m}\right) = \sum_{a \in B \oplus C} c(n - 1, a), \quad (1 \leq m \leq n), \qquad (26)$$

and hence the normalized total Banzhaf power is given

$$NTBP(X_m) = \frac{1}{n}, \quad (1 \leq m \leq n), \qquad (27)$$

as expected. In retrospect, we note that we might not have really needed to carry out the aforementioned detailed calculations, because we could have deduced directly from the symmetry of the voting system that the voting powers of the voters are going to be equal. Though the computations of this section are not particularly useful for their own sake, they are potentially of notable benefit in handling certain voting systems that possess dominant partial symmetries among voters, which is the case for many notable voting systems, including each of the two voting systems in Section 4.

## 4. Examples of Weighted Voting Systems

### 4.1. The European Economic Community ($EEC$)

The European Economic Community ($EEC$) is the first ancestor (that lasted from 1958 to 1973) of the present-day European Union ($EU$). It is a weighted voting system of six members, which is described by a threshold $T = 12$, and a vector of six weights

$$\mathbf{W} = [W_F \quad W_G \quad W_I \quad W_B \quad W_N \quad W_L]^T = [\ 4 \quad 4 \quad 4 \quad 2 \quad 2 \quad 1\ ]^T, \quad (28)$$

where the subscripts F, G, I, B, N, and L respectively denote the west European countries of France, Germany (then West Germany), Italy, Belgium, the Netherlands and Luxembourg [7]. The system is described by a threshold switching function, whose minimal or complete sum is [7]

$$f(F, G, I, B, N, L) = FGI \lor FGBN \lor FIBN \lor GIBN. \quad (29)$$

The function $f(F, G, I, B, N, L)$ is independent of (vacuous in) the variable $L$. This function is not a genuine 6-variable function as it degenerates into a 5-variable one. Luxembourg is, in fact, a dummy voter within the $EEC$ system, such that $\frac{\partial f}{\partial L} = 0$ and subsequently $TBP(L) = 0$. This finding does not follow immediately from a hasty, superficial and cursory inspection of the weights in (28).

The function $f(F, G, I, B, N, L)$ can be rewritten in terms of symmetric switching functions as

$$f(F, G, I, B, N, L) = Sy\,(3; \{3\}; F, G, I) \lor Sy\,(3; \{2,3\}; F, G, I)\, BN. \quad (30)$$

We convert this expression into a disjoint sop one by multiplying the second term by the complement of the first term (according to the Reflection Law [24, 32]), namely

$$f(F, G, I, B, N, L) = Sy\,(3; \{3\}; F, G, I) \lor Sy\,(3; \{0,1,2\}; F, G, I)\, Sy\,(3; \{2,3\}; F, G, I)\, BN$$

$$= Sy\,(3; \{3\}; F, G, I) \lor Sy\,(3; \{2\}; F, G, I)\, BN. \quad (31)$$

Since the two terms in (30) above are now disjoint, we can replace the OR operator ($\lor$) by an XOR operator ($\oplus$), namely

$$f(F, G, I, B, N, L) = Sy\,(3; \{3\}; F, G, I) \oplus Sy\,(3; \{2\}; F, G, I)\, BN. \quad (32)$$

Due to partial symmetries, we note that $TBP(F) = TBP(G) = TBP(I)$, and $TBP(N) = TBP(B)$. Hence, it suffices to compute the Boolean derivative w.r.t. one of the variables $F, G,$ and $I$ (say $F$), and one of the variables $B$ and $N$ (say $B$), namely

$$\frac{\partial f}{\partial F} = (Sy\,(2; \boldsymbol{\phi};\, G, I) \oplus Sy\,(2; \{2\}; G, I)) \oplus (Sy\,(2; \{2\}; G, I) \oplus Sy\,(2; \{1\}; G, I))\ BN.$$

$$= Sy\,(2; \{2\}; G, I)\,(1 \oplus BN) \oplus Sy\,(2; \{1\}; G, I)\, BN. \quad (33)$$

$$\frac{\partial f}{\partial B} = f(X) = Sy\,(3;\{2\};F,G,I)\ N. \tag{34}$$

which correspond to total Banzhaf powers of

$$TBP(F) = wt\left(\frac{\partial f}{\partial F}\right) = (1)(4-1) + (2)(1) = 5, \tag{35}$$

$$TBP(B) = wt\left(\frac{\partial f}{\partial B}\right) = (3)(1) = 3. \tag{36}$$

Finally, the vectors of total Banzhaf powers and normalized total Banzhaf powers are

$$\boldsymbol{TBP} = [\,5\quad 5\quad 5\quad 3\quad 3\quad 0\,]^T,\ \boldsymbol{NTBP} = \left[\tfrac{5}{21}\quad \tfrac{5}{21}\quad \tfrac{5}{21}\quad \tfrac{3}{21}\quad \tfrac{3}{21}\quad 0\right]^T. \tag{37}$$

**4.2. The Extended European Economic Community ($EEEC$)**

The Extended European Economic Community ($EEEC$) is again a predecessor of the contemporary European Union ($EU$) [7]. This 9-member weighted voting system emerged in 1973 when the $EEC$ was extended through the addition of three new member countries, which are: the United Kingdom of Britain (R), Denmark (D) and Ireland (E). The weight vector was updated for this new system to become:

$$\boldsymbol{W'} = [W'_F\quad W'_G\quad W'_I\quad W'_R\quad W'_B\quad W'_N\quad W'_D\quad W'_E\quad W'_L]^T$$
$$= [10\quad 10\quad 10\quad 10\quad 5\quad 5\quad 3\quad 3\quad 2\,]^T, \tag{38}$$

while the threshold was reset to $T' = 41$. The system is described by a threshold switching function, whose minimal or complete sum is [7]

$$\begin{aligned}
f(F,G,I,R,B,N,D,E,L) \\
&= (BNL \vee BNE \vee BND \vee NED \vee BED)\,(FGI \vee FGR \vee FIR \vee GIR) \\
&\quad \vee (B \vee N \vee D \vee E \vee L)\,FGIR. \\
&= (BNL \vee BNE \vee BND \vee NDE \vee BDE)\,Sy(4;\{3,4\};F,G,I,R) \\
&\quad \vee (B \vee N \vee D \vee E \vee L)\,Sy(4;\{4\};F,G,I,R) \\
&= (BNL \vee BNE \vee BND \vee NDE \vee BDE)\,(Sy(4;\{3\};F,G,I,R) \vee Sy(4;\{4\};F,G,I,R)) \\
&\quad \vee (B \vee N \vee D \vee E \vee L)\,Sy(4;\{4\};F,G,I,R). 
\end{aligned} \tag{39}$$

The function $f(F,G,I,R,B,N,D,E,L)$ is a genuine function of its nine arguments, and hence, none of the voters it represents is dummy. Noting that

$$(BNL \vee BNE \vee BND \vee NDE \vee BDE) \le (B \vee N \vee D \vee E \vee L), \tag{40}$$

we reduce the expression of $f(F,G,I,R,B,N,D,E,L)$ in (39) to

$$\begin{aligned}
f(F,G,I,R,B,N,D,E,L) &= (BNL \vee BNE \vee BND \vee NDE \vee BDE)\,Sy(4;\{3\};F,G,I,R) \\
&\quad \vee (B \vee N \vee D \vee E \vee L)\,Sy(4;\{4\};F,G,I,R).
\end{aligned} \tag{41}$$

The two terms in the R.H.S. of (41) are mutually disjoint, and it is immaterial to separate them with an OR or an XOR. Furthermore, we employ disjointing techniques to replace other ORs by XORs and hence we rewrite (41) as the following easy-to-differentiate expression

$$f(F,G,I,R,B,N,D,E,L)$$
$$= (BNL \oplus BNE\bar{L} \oplus BND\bar{E}\,\bar{L} \oplus \bar{B}NDE$$
$$\oplus BDE\bar{N})\ Sy(4;\{3\};F,G,I,R)$$
$$\oplus (1 \oplus \bar{B}\,\bar{N}\,\bar{D}\,\bar{E}\,\bar{L})\ Sy(4;\{4\};F,G,I,R). \tag{42}$$

Due to partial symmetries, we note that $TBP(F) = TBP(G) = TBP(I) = TBP(R)$, $TBP(B) = TBP(N)$, and $TBP(D) = TBP(E)$. Hence, it suffices to compute the Boolean derivative w.r.t. one of the variables $F, G, I$ and $R$ (say $F$), one of the variables $B$ and $N$ (say $B$), one of the variables $D$ and $E$ (say $D$), and $L$, namely

$$\frac{\partial f}{\partial F} = (BNL \oplus BNE\bar{L} \oplus BND\bar{E}\,\bar{L} \oplus \bar{B}NDE \oplus$$
$$BDE\bar{N})\ (Sy(3;\{3\};G,I,R) \oplus Sy(3;\{2\};G,I,R) \oplus (1 \oplus \bar{B}\,\bar{N}\,\bar{D}\,\bar{E}\,\bar{L})\ Sy(3;\{3\};G,I,R)$$

$$= (1 \oplus \bar{B}\,\bar{N}\,\bar{D}\,\bar{E}\,\bar{L} \oplus BNL \oplus BNE\bar{L} \oplus BND\bar{E}\,\bar{L} \oplus \bar{B}NDE$$
$$\oplus BDE\bar{N})\ Sy(3;\{3\};G,I,R)$$
$$\oplus (BNL \oplus BNE\bar{L} \oplus BND\bar{E}\,\bar{L} \oplus \bar{B}NDE \oplus BDE\bar{N})\ Sy(3;\{2\};G,I,R). \tag{43}$$

$$\frac{\partial f}{\partial B} = (NL \oplus NE\bar{L} \oplus ND\bar{E}\,\bar{L} \oplus NDE \oplus DE\bar{N})\ Sy(4;\{3\};F,G,I,R)$$
$$\oplus \bar{N}\,\bar{D}\,\bar{E}\,\bar{L}\ Sy(4;\{4\};F,G,I,R). \tag{44}$$

$$\frac{\partial f}{\partial D} = (BN\bar{E}\,\bar{L} \oplus \bar{B}NE \oplus BE\bar{N})\ Sy(4;\{3\};F,G,I,R)$$
$$\oplus \bar{B}\,\bar{N}\,\bar{E}\,\bar{L}\ Sy(4;\{4\};F,G,I,R). \tag{45}$$

$$\frac{\partial f}{\partial L} = (BN \oplus BNE \oplus BND\bar{E})\ Sy(4;\{3\};F,G,I,R) \oplus \bar{B}\,\bar{N}\,\bar{D}\,\bar{E}\ Sy(4;\{4\};F,G,I,R)$$
$$= BN\bar{D}\,\bar{E}\ Sy(4;\{3\};F,G,I,R) \oplus \bar{B}\,\bar{N}\,\bar{D}\,\bar{E}\ Sy(4;\{4\};F,G,I,R). \tag{46}$$

which correspond to total Banzhaf powers of

$$TBP(F) = wt\left(\frac{\partial f}{\partial F}\right) = (32 - (1 + 4 + 2 + 1 + 2 + 2))(1) + (4 + 2 + 1 + 2 + 2)(3) =$$
$$20 + 33 = 53, \tag{47}$$

$$TBP(B) = wt\left(\frac{\partial f}{\partial B}\right) = (7)(4) + (1)(1) = 29. \tag{48}$$

$$TBP(D) = wt\left(\frac{\partial f}{\partial D}\right) = (1 + 2 + 2)(4) + (1)(1) = 21. \tag{49}$$

$$TBP(L) = wt\left(\frac{\partial f}{\partial L}\right) = (1)(4) + (1)(1) = 5. \tag{50}$$

where we obtained the weight of the leading function in (44) by expanding this function into 4 subfunctions with easy-to-compute weights that add to the weight of the parent function, namely

$$wt(NL \oplus NE\bar{L} \oplus NDE\,\bar{L} \oplus NDE \oplus DE\bar{N}) = wt(NL \oplus NE\bar{L} \oplus NDE\,\bar{L} \oplus DE) =$$
$$wt(NL) + wt(NL \oplus N\bar{L}) + wt(NL \oplus N\bar{L}) + wt(NL \oplus N\bar{L} \oplus 1) = 1 + 2 + 2 + 2 = 7. \quad (51)$$

Finally, the vectors of total Banzhaf powers and normalized total Banzhaf powers are

$$\boldsymbol{TBP} = [\,53 \quad 53 \quad 53 \quad 53 \quad 29 \quad 29 \quad 21 \quad 21 \quad 5\,]^T, \quad (52)$$

$$\boldsymbol{NTBP} = \left[\frac{53}{317} \quad \frac{53}{317} \quad \frac{53}{317} \quad \frac{53}{317} \quad \frac{29}{317} \quad \frac{29}{317} \quad \frac{21}{317} \quad \frac{21}{317} \quad \frac{5}{317}\right]^T. \quad (53)$$

## 5. Conclusions

This paper gives a brief overview of switching-algebraic techniques for computing the Banzhaf voting power. The paper focuses on comprehending the inherent properties of a voting system by analyzing outcomes through the voting power approach, and is hence capable of ferreting out hidden facts that are not otherwise self-evident The paper also attempts to make the most of symmetry considerations, which are typically present in many voting systems of practical importance. Moreover, the paper also offers a useful tutorial coverage of the subject matter of weighted voting systems, and it translates many concepts of this subject matter to the switching-algebraic domain.

In the foregoing computation, we made the explicit assumption that variables in the considered systems are statistically independent. For future work, we need to relax this assumption, and to consider the issues of alliances and partisan identification and commitment, which leads to similar voting patterns among many voters (analogous to common-cause effect in reliability studies).

For more future work, we plan to employ switching-algebraic techniques to tackle other standard voting systems such as the voting system of the United Nations security Council [35] and the vector-weighted 537-variable system that describes the federal voting system of the United States of America (the system comprising the President and Vice-President of the USA plus the Congress (the Senate and the House of Representatives) [2]. We are going also to explore some of the paradoxes associated with voting powers, such as the paradox of redistribution, the paradox of new members, the quarrelling paradox, the donation paradox, and the paradox of large size [36-42]. The power indices are utilized herein in a descriptive sense, but could be otherwise used in a normative sense, which gives rise to a design or inverse problem that deals with the allocation of power to the voters according to the pre-established target [42-47].

## Appendix A: The Boolean Difference or Derivative

A switching function (a two-valued Boolean function) on $n$ variables is a mapping from $B_2^n = \{0,1\}^n$ into $B_2 = \{0,1\}$, that is denoted by $f(\boldsymbol{X}) = f(X_1, X_2, \cdots, X_i, \ldots, X_{n-1}, X_n)$. The partial derivative (or Boolean difference) of $f(\boldsymbol{X})$ w.r.t. $X_i$ ($1 \leq i \leq n$) is [6, 7, 25, 26, 34]

$$\frac{\partial f}{\partial X_i} = f(X_1, X_2, \cdots, \bar{X}_i, \ldots, X_{n-1}, X_n) \oplus f(X_1, X_2, \cdots, X_i, \ldots, X_{n-1}, X_n). \quad (A.1)$$

This can be seen to be equivalent to

$$\frac{\partial f}{\partial X_i} = f(X|X_i = 0) \oplus f(X|X_i = 1). \quad (A.2)$$

where

$$f(X|X_i = 0) = f(X_1, X_2, \cdots, 0, \ldots, X_{n-1}, X_n) = f(X) / \overline{X}_i, \quad (A.3a)$$

$$f(X|X_i = 1) = f(X_1, X_2, \cdots, 1, \ldots, X_{n-1}, X_n) = f(X) / X_i, \quad (A.3b)$$

are called subfunctions, restrictions or quotients (ratios) of $f(X)$. Their Karnaugh maps are obtained by *splitting* the Karnaugh map of $f(X)$ into two halves, *viz.*, the asserted domains for $\overline{X}_i$ and $X_i$. The Boolean difference is then obtained by *folding* one of these two halves onto the other and performing *XORing* cell-wise. Some of the important properties of the Boolean difference are (for $A$ and $B$ being independent of $X_i$)

$$\{f(X) = f_1(X) \oplus f_2(X)\} \rightarrow \{\frac{\partial f(X)}{\partial X_i} = \frac{\partial f_1(X)}{\partial X_i} \oplus \frac{\partial f_2(X)}{\partial X_i}\}, \quad (A.4)$$

$$\{f(X) = f_1(X) \vee f_2(X)\} \rightarrow \{\frac{\partial f(X)}{\partial X_i} = \overline{f}_1(X) \frac{\partial f_2(X)}{\partial X_i} \oplus \frac{\partial f_1(X)}{\partial X_i} \overline{f}_2(X) \oplus \frac{\partial f_1(X)}{\partial X_i} \frac{\partial f_2(X)}{\partial X_i}\}, \quad (A.5)$$

$$\frac{\partial f}{\partial X_i} = \frac{\partial f}{\partial \overline{X}_i} = \frac{\partial \overline{f}}{\partial X_i} = \frac{\partial \overline{f}}{\partial \overline{X}_i}, \quad (A.6)$$

$$\frac{\partial (AX_i)}{\partial X_i} = \frac{\partial (AX_i)}{\partial \overline{X}_i} = A, \quad (A.7)$$

$$\frac{\partial (B)}{\partial X_i} = \frac{\partial (B)}{\partial \overline{X}_i} = 0. \quad (A.8)$$

Equation (A.4) indicates that the partial differentiation operator ($\frac{\partial}{\partial X_i}$) commutes with the XOR operator (). By contrast, the partial differentiation operator does not commute with the OR operator, and a quite involved formula is needed (A.5) to differentiate an ORed expression. That is the reason why we prefer to pre-process a switching function before differentiating it by converting it first to a disjoint sum-of-products form and then replacing OR operators by XOR ones.

## Appendix B: Computing the Weight of a Switching Function

We interpret a switching function $f(X) = f(X_1, X_2, \cdots, X_{n-1}, X_n)$ as the output column $f$ of its truth table, which is a binary vector of length $2^n$, namely

$$f = [f(0,0,\ldots 0,0) \quad f(0,0,\ldots,0,1) \quad f(0,0,\ldots,1,0) \ldots f(1,1,\ldots,1,1)]^T. \quad (B.1)$$

The weight $wt(f)$ of the switching function $f(X)$ is then defined as the number of ones in this truth-table vector $f$. This weight is naturally bounded by $0 \leq wt(f) \leq 2^n$. If the weight is normalized by $2^n$, then it is called the syndrome $s(f)$ of the switching function $f(X)$, and is bounded by $0 \leq s(f) \leq 1$. The syndrome might be interpreted as a probability, and it serves as the first of the $2^n$ spectral coefficients of $f(X)$.

The real transform $R(\boldsymbol{p}) = R(p_1, p_2, \cdots, p_n)$ of a switching function $f(X)$, referred to by $R\ell(f)$, is defined to enjoy two characterictics [32, 48-51], namely:

a) $R(\boldsymbol{p})$ is a multi-affine continuous real function of $n$ continuous real variables $\boldsymbol{p} = [p_1\ p_2\ \cdots\ p_n]^T$. If all arguments other than argument $p_i$ ($1 \leq i \leq n$) are kept constant, then $R(\boldsymbol{p})$ takes the form $(A_i + B_i\ p_i)$ (with $A_i$ and $B_i$ being constants), i.e., it reduces to a straight-line relation or a first-degree polynomial in the argument $p_i$.

b. $R(\boldsymbol{p})$ shares the same "truth table" with $f(X)$, i.e.

$$R(\boldsymbol{p} = \boldsymbol{t}_j) = f(X = \boldsymbol{t}_j), \qquad \text{for } j = 0, 1, \ldots, (2^n - 1), \qquad (B.2)$$

where $\boldsymbol{t}_j$ is the jth input line of the truth table; $\boldsymbol{t}_j$ is an $n$-vector of binary components such that

$$\sum_{i=1}^{n} 2^{n-i}\ t_{ji} = j, \qquad \text{for } j = 0, 1, \ldots, (2^n - 1). \qquad (B.3)$$

We emphasize that characteristic (b) above is not sufficient by itself to produce a unique $R(\boldsymbol{p})$ unless it is augmented by the requirement (a) that $R(\boldsymbol{p})$ be multi-affine [32]. If both the real transform $R$ and its arguments $\boldsymbol{p}$ are restricted to binary values (i.e., if $R: \{0, 1\}^n \to \{0, 1\}$) then $R$ becomes the multilinear form of a switching function studied extensively by Schneeweiss [30, 31], typically used to mimic the structure function [20, 52, 53] in engineering study of system reliability.

The real transform $R(\boldsymbol{p})$ might be viewed as a mapping $R(\boldsymbol{p}): \boldsymbol{R}^n \to \boldsymbol{R}$, where $\boldsymbol{R}$ is the entire real line. This transform is also named the probability transform, a name which stems from the fact that the mapping might be recast correctly as $R(\boldsymbol{p}): [0.0, 1.0]^n \to [0.0, 1.0]$, and hence both $R$ and $\boldsymbol{p}$ could be interpreted as probabilities.

The following paragraph highlights a convenient way for obtaining the real transform of a switching function $f(X)$ by first expressing it in a disjoint sum-of-products (s-o-p) form [32]

$$f(X) = \bigvee_{k=1}^{m} D_k, \qquad (B.4)$$

where

$$D_i \wedge D_j = 0, \qquad \forall\ i, j, \qquad (B.5)$$

$$D_k = \left(\bigwedge_{i \in I_{k_1}} X_i\right)\left(\bigwedge_{i \in I_{k_2}} \overline{X}_i\right), \qquad \forall\ k. \qquad (B.6)$$

Here, $I_{k_1}$ and $I_{k_2}$ are the sets of indices for uncomplemented literals and complemented literals in the product $D_k$. None of these literals is redundant, for otherwise, redundancy of a literal is eliminable through idempotency of AND ($X_i \wedge X_i = X_i$, $\overline{X}_i \wedge \overline{X}_i = \overline{X}_i$). The real transform $R(\boldsymbol{p}) = R\ell(f)$ is given by

$$R(\boldsymbol{p}) = R\ell(f) = \sum_{k=1}^{m} T\{D_k\}(\boldsymbol{p}), \qquad (B.7)$$

where

$$T\{D_k\}(\mathbf{p}) = \left(\prod_{i \in I_{k_1}} p_i\right) \left(\prod_{i \in I_{k_2}} (1 - p_i)\right), \quad \forall k, \tag{B.8}$$

The R.H.S. of (B.7) is obtained from that of (B.4) by replacing the Boolean *AND* operator by the real multiplication operator, the Boolean *OR* operator by the real addition operator, each un-complemented Boolean variable $X_i$ by the real variable $p_i$, and each complemented Boolean variable $\overline{X}_i$ by the real variable $(1 - p_i)$.

Once the real transform $R(\mathbf{p})$ of the switching function $f(X)$ is obtained, then its weight is readily expressed as

$$wt(f) = 2^n * R(\mathbf{2^{-1}}) = 2^n * R(2^{-1}, 2^{-1}, \cdots, 2^{-1}), \tag{B.9}$$

where $\mathbf{2^{-1}}$ means a vector of $n$ elements, each of which is $2^{-1} = 0.5$. Furthermore, if $f(X)$ is expressed by the disjoint s-o-p form (B.4), then its weight is given by

$$wt(f) = \sum_{k=1}^{m} wt(D_k) = 2^n * \sum_{k=1}^{m} T\{D_k\}(\mathbf{2^{-1}}) = \sum_{k=1}^{m} 2^{(n-\ell(D_k))}, \tag{B.10}$$

where $\ell(D_k)$ is the number of irredundant literals in the product $D_k$, e. g., $\ell(1) = 0$, $\ell(X_i) = \ell(\overline{X}_i) = 1$, $\ell(X_i X_j) = \ell(X_i \overline{X}_j) = 2$. The logical value 0 is the identity of the *ORing* operation, and is not viewed as a logical product $D_k$ at all. For convenience, we take $\ell(0) = \infty$, so were we to have a product $D_k = 0$, we ensure that its weight is $wt(D_k) = 0$. The Minterm Canonical Form of $f(X)$ is a special case of the disjoint expansion (B.4), for which $m$ depicts the number of minterms or the number of true configurations of $f(X)$. Here $\ell(D_k) = n$, $\forall k$, and (B.10) produces the correct result $wt(f) = m$ in this case.

If the function $f(X)$ is not available in the disjoint s-o-p form (B.4), but in a general s-o-p form that is not necessarily disjoint,

$$f(X) = \bigvee_{i=1}^{n_p} P_i, \tag{B.11}$$

then the weight of $f(X)$ is given by an appropriate version of the Inclusion-Exclusion (IE) Principle [32, 54-55] as follows

$$wt(f) = \sum_{i=1}^{n_p} wt\{P_i\} - \sum\sum_{1 \leq i < j \leq n_p} wt\{P_i \wedge P_j\} + \sum\sum\sum_{1 \leq i < j < k \leq n_p} wt\{P_i \wedge P_j \wedge P_k\} - \cdots + (-1)^{n_p - 1} wt\{\bigwedge_{i=1}^{n_p} P_i\}. \tag{B.12}$$

where the weight of a product is (according to (B.10)) equal to 2 raised to the power of the total number of variables minus the number of irredundant literals in the product.

If the function $f(X)$ is a symmetric switching function $Sy(n; A; X)$, then its weight can be obtained by summing the combinatorial (binomial) coefficients $n$ choose $a$, denoted $c(n, a)$, for all integers $a$ that belong to the characteristic set $A$, namely:

$$wt(Sy(n; A; X)) = \sum_{a \in A} c(n, a). \tag{B.13}$$

If a function $f(X, Y)$ is a conjunction of two functions $f_1(X)$ and $f_2(Y)$, where $X$ and $Y$ are non-overlapping sets of arguments, then the weight of $f(X, Y)$ is the arithmetic product of the weights of $f_1(X)$ and $f_2(Y)$.

$$\{f(X,Y) = f_1(X) \wedge f_2(Y)\} \rightarrow \{wt(f(X,Y)) = wt(f_1(X)) * wt(f_2(Y))\}. \quad (B.14)$$

If a function $f(X)$ is a disjunction (or XORing) of two disjoint functions $f_1(X)$ and $f_2(X)$, then its weight is the arithmetic sum of their weights

$$\{f(X) = f_1(X) \vee f_2(X) = f_1(X) \oplus f_2(X), \quad f_1(X) \wedge f_2(X) = 0\} \rightarrow$$
$$\{wt(f(X)) = wt(f_1(X)) + wt(f_2(X))\}. \quad (B.15)$$

Moreover, if a switching function is expanded about $m$ of its arguments into $2^m$ subfunctions (which are naturally disjoint), then the weight of this parent function is the sum of the weights of these subfunctions. The weights of a function (of $n$ arguments) and its complement add to $2^n$

$$wt\left(\overline{f}(X)\right) = wt(1 \oplus f(X)) = 2^n - wt(f(X)). \quad (B.16)$$

**Example B.1:**

The 2-out-of-3 function

$$f(X) = f(X_1, X_2, X_3) = X_1 X_2 \vee X_2 X_3 \vee X_1 X_3 = Sy(3; \{2,3\}; X_1, X_2, X_3), \quad (B.17)$$

is represented by the Karnaugh map in Fig. B.1. The function is covered with non-overlapping loops, and hence it is expressed by the disjoint s-o-p expression

$$f(X_1, X_2, X_3) = X_1 X_2 \vee \overline{X}_1 X_2 X_3 \vee X_1 \overline{X}_2 X_3. \quad (B.18)$$

Therefore, its weight is obtained correctly via (B.10) as

$$wt(f) = 2^{3-2} + 2^{3-3} + 2^{3-3} = 2 + 1 + 1 = 4. \quad (B.19)$$

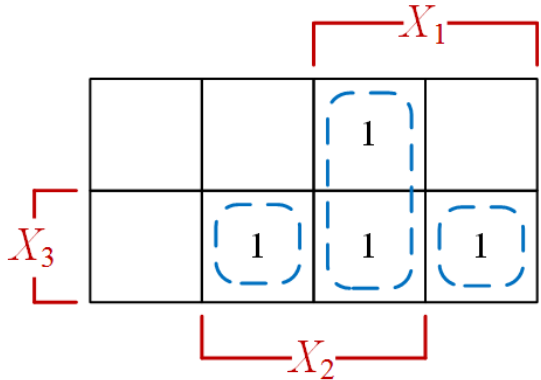

Fig. B.1. A Karnaugh map representing a 2-out-of-3 function with disjoint coverage. As usual, 0 entries within the map are left blank.

This weight might also be computed via the IE principle (B.12) as

$$wt(f) = wt(X_1X_2) + wt(X_2X_3) + wt(X_1X_3) - wt(X_1X_2 \wedge X_2X_3) - wt(X_1X_2 \wedge X_1X_3)$$
$$- wt(X_2X_3 \wedge X_1X_3) + wt(X_1X_2 \wedge X_2X_3 \wedge X_1X_3)$$

$$= wt(X_1X_2) + wt(X_2X_3) + wt(X_1X_3) - wt(X_1X_2X_3) - wt(X_1X_2X_3) - wt(X_1X_2X_3)$$
$$+ wt(X_1X_2X_3) = 2^{3-2} + 2^{3-2} + 2^{3-2} - 2^{3-3} - 2^{3-3} - 2^{3-3} + 2^{3-3}$$

$$= 2 + 2 + 2 - 1 - 1 - 1 + 1 = 4. \tag{B.20}$$

Finally, we can recognize the symmetry of the function $f(X) = y(n; A; X)$, with $n = 3$ and $A = \{2,3\}$, and then employ (B.13) to obtain

$$wt(f) = c(3,2) + c(3,3) = 3 + 1 = 4. \tag{B.21}$$

For the characteristic set $A$, the corresponding sets in (23) are $B = \{2\}, C = \{1, 2\}$, and $B \oplus C = \{1\}$, and hence the total Banzhaf power of any of the arguments is given by

$$TBP(X_m) = wt\left(\frac{\partial Sy(n;\{2,3\};X)}{\partial X_m}\right) = c(2,1) = 2, \ (1 \leq m \leq 3). \tag{B.22}$$